\documentclass[12pt,preprint]{aastex}

\begin{document}

\title{Fast transition between Classical and Weak lined T Tauri stars due to external UV dissipation}
\author{Jin Zeng Li\altaffilmark{1}, Travis A.
Rector\altaffilmark{2}}
\altaffiltext{1}{National Astronomical
Observatories, Chinese Academy of Sciences, Beijing 100012, China;
ljz@bao.ac.cn}
\altaffiltext{2}{University of Alaska at Anchorage,
3211 Providence Drive, Anchorage, AK 99508}

\begin{abstract}

The discovery of optical jets immersed in the strong UV radiation
field of the Rosette Nebula sheds new light on, but meanwhile
poses challenges to, the study of externally irradiated jets.  The
jet systems in the Rosette are found to have a high state of
ionization and show unique features.  In this paper, we
investigate the evolutionary status of the jet driving sources for
young solar-like stars.  To our surprise, these jet sources
indicate unexpected near infrared properties with no excess
emission. They are bathed in harsh external UV radiation such that
evaporation leads to a fast dissipation of their circumstellar
material.  This could represent a transient phase of evolution of
young solar-like stars between classical and weak lined T Tauri
stars.  Naked T Tauri stars formed in this way have
indistinguishable evolutionary ages from those of classical T
Tauri stars resulting from the same episode of star formation.
However, it would be hard for such sources to be identified if
they are not driving an irradiated jet in a photoionized medium.

\end{abstract}


\section{Introduction}

One of the major advancements in recent years in the study of
Herbig-Haro (HH) flows has been the discovery of externally
irradiated jets within the confines of HII regions (Reipurth et
al. 1998).  Similar jets immersed in photoionized medium were
identified in the Orion nebula and near the late B stars in NGC
1333 (Bally et al. 2000; Bally \& Reipurth 2001).

The Rosette Nebula is a spectacular HII region that has been
excavated by strong winds from the OB stars at the center of the
young open cluster NGC 2244, recently identified as a possible
twin cluster based upon near-infrared studies from the 2MASS
survey (Li 2005). This emerging young open cluster is found to
have a turn-off age of about 1.9 x 10 $^6$ yrs (Park \& Sung
2002). A set of primarily two jet systems with an externally
irradiated origin, HH 889 \& HH 890 (please refer to
http://casa.colorado.edu/hhcat/) were discovered in the Rosette
Nebula. Note that these two HH jets were formerly called the
Rosette HH1 (Li \& Rector, 2004; Li, Chu \& Gruendl 2006) \& HH2
(Li 2003), respectively.

The HH jets discovered in Rosette (Li \& Rector, 2004; Li 2003)
and their only rivals in the close vicinity of $\sigma$ Orionis
(Reipurth et al. 1998) are found to share many similar properties,
which are commensurate with an irradiated origin of the disk-jets
systems: (i) all the jet sources are identified in HII regions,
bathed in harsh UV ionization from nearby (within a few parsecs)
massive OB stars which dominate the excavation of the HII regions.
Contrary to conventional jet exciting sources, they show very low
extinction along the line of sight and are therefore optically
visible, with spectra characteristic of T Tauri stars; (ii) none
is detected by IRAS, indicating the lack of circumstellar
materials in the form of extended disks and/or envelopes; (iii)
all show a normal shock-excited origin of [SII] at the base of the
jets, which declines rapidly with distance from the driving source
and turns to be H$\alpha$ dominated, indicating a photoionized
nature; (iv) all have a highly asymmetric or even unipolar
morphology, indicating perhaps much different jet forming
conditions in the launch and collimation regions. As a comparison,
jets found in the outskirts of the Orion Nebula show a bipolar but
'C' shaped structure, bending away from the nebula core (Bally et
al. 2000).

However, the Rosette HH jets show additional features not seen in
the $\sigma$ Orionis jets.  For example, both the HH 889 \& 890
jets show a high excitation origin, due to perhaps the more
intense extreme ultraviolet (EUV) radiation from the exciting OB
stars in Rosette, which is 1-2 orders of magnitudes higher than
the ionizing fields in orion (Reipurth et al. 1998; Bally et al.
2000). Bally et al. (1998) also found shock structures indicating
significant [OIII] emission in the immediate vicinity of the
Trapezium stars (within 30${"}$ and at a projected distance of
0.06 pc from $\theta$ Ori C if a distance of 430 pc to the Orion
Nebula is adopted), which suggests a shock excited origin of the
associated materials and a dependence of the excitation on the
intensity of EUV ionization and impacts of fast stellar winds, and
hence the distance from the intervening hot OB star. On the other
hand, the Rosette jets are not beaming away from any specific OB
star or a group of nearby exciting stars in projection.
Furthermore, the HH 889 source show in its spectra significant
signatures of mass accretion or inflow but no detectable veiling.
Successful interpretation of these special features may lead to
the final solution of jet formation associated with these and
other young stellar objects (YSOs).

Li \& Rector (2004) proposed that the discovery of the Rosette
jets provides evidence of efficient dissipation of circumstellar
disks and envelopes in the vicinity of massive OB stars. This
provides indirect evidence for the formation of isolated brown
dwarfs (BDs) and free-floating giant planets.  UV dissipation of
pre-existing protostellar systems in the Rosette by nearby OB
stars would prevent these objects from evolving into low-mass
stars. Such a mechanism has been shown to be possible and
effective by the theoretical studies of Whitworth \& Zinnecker
(2004).

It is therefore important to explore the nature of jet formation
and disk dissipation of low-mass YSOs in the close vicinity of
ionizing sources such as OB stars, as the occurrence of such OB
clusters and associations is common in the Galaxy.  And our solar
system may have formed in such an environment (Looney, Tobin \&
Fields 2006). In addition, there has been an unresolved debate on
whether there is an evolutionary sequence from classical T Tauri
stars (CTTS) to weak lined T Tauri stars (WTTS) as the
circumstellar materials around a protostar are gradually
exhausted, or they could otherwise be born different in the same
regions of active star formation. The latter will require a quick
disk dissipation due to external forces after the primordial stage
of protostellar collapse and the formation of a protostar. A
detailed study of the Rosette jet systems may provide strong test
on an abrupt evolution of CTTS to WTTS due to external
photoionization of their protostellar disks in massive star
forming regions.  WTTS formed in this way have indistinguishable
evolutionary ages as those of CTTS originated from the same
episode of star formation but may indicate substantially different
observational properties as will be discussed in this paper.

\section{Observations and data reduction}

Spectroscopic observations of the HH 889 source were carried out
with the 2.16 m telescope of the National Astronomical Observatory
of the Chinese Academy of Sciences (NAOC) on Jan. 5, 2005.  An OMR
(Optomechanics Rsearch Inc.) spectrograph and a Tecktronix 1024 x
1024 CCD were used.  A 100~\AA~mm$^{-1}$ grating and the 2 $"$
slit resulted in a two-pixel resolution of the spectra of 4.8~\AA.
The spectral resolution of these observations are therefore not
significantly better than those employed in the discover of the HH
889 jet (Li \& Rector 2004). However, it is suitable for the faint
exciting source of the jet in isolating H$\alpha$ from the nearby
[NII] emission lines.

Two consecutive spectra of the HH 890 source were achieved with
the Beijing Faint Object Spectrograph and Camera (BFOSC) and a
thinned back-illuminated Orbit 2k$\times$2k CCD equipped to the
NAOC 2.16m telescope on Jan. 4, 2005. Due to the faintness of the
HH 890 source with USNO magnitudes of 18.7 in R and 18.2 in B,
respectively, the G4 grating is employed, which gives a two-pixel
resolution of 8.3~\AA.

The spectroscopic data were reduced following standard procedures
and packages in IRAF. The CCD reduction included bias and
flat-field correction, successful nebular background subtraction,
and cosmic rays removal. Wavelength calibration was performed
based on helium-argon lamps exposed at both the beginning and the
end of the observations every night. Flux calibration of each
spectrum was conducted based on observations of at least 2 of the
KPNO spectral standards (Massey et al. 1988) per night.

\section{Low-resolution spectra of the jet sources in Rosette}

Fig. 1 presents the spectrum of the HH 889 source around
H$\alpha$. At a resolution of $\sim$4.8\AA, the H$\alpha$ emission
is successfully resolved from the nearby [NII] emission lines.
This is important as the HH 889 source was found to have
complicated profile of H$\alpha$ (Li \& Rector 2004). The
forbidden emission lines of [NII] indicate a line ratio of
I($\lambda$6583)/I($\lambda$6548)=3.0, commensurate with their
origin from photoionized medium. The H$\alpha$ emission, however,
shows again an Inverse P-Cygni profile as presented by Li \&
Rector (2004) with a lower spectral resolution. A red-displaced
absorption feature is clearly resolved, which shows a receding
velocity of $\sim$ 150 km s$^{-1}$ associated with likely mass
accretion or inflow.  The weak H$\alpha$ in emission from the
source gives an equivalent width of only 1.7~\AA. This leads to
the classification of a WTTS nature of the source based on
predominantly the intensity of its H$\alpha$ emission. This issue
will be further elaborated in Section 5.

Here we discuss in detail possible interpretations to the
red-displaced absorption associated with H$\alpha$: (1) It is due
to a continuous accretion or rather inflow of ablated disk
materials. However, it is essentially different from conventional
IPC profiles as signatures of enhanced accretion onto the central
protostar. As noted by Li \& Rector (2004) and convinced by our
time series low-resolution spectroscopy, no noticeable veiling in
the continuum emission was detected. Could this imply that the
accretion flow covers most of the projected surface area of the
protostar in the region where most of the absorption is occurring,
possibly many stellar radii in front of the star? This, on the
other hand, may be closely related to how the jet is sustained in
the lack of circumstellar materials as stated below, and poses
challenge to traditional knowledge of jet formation and especially
mass loading. In this case, EUV ionziation of the relic disk
material may have provided an efficient mechanism for mass
ejection via a highly deformed star-disk magnetosphere facing the
strong radiation field. This could explain why the collimated jet
survives its fierce environment. It, however, leads to a fast
dissipation of the relic disk through both mass ejection and
photoevaporation. (2) Alternatively, the absorption line could be
due to absorption and obscuration of background nebular emission
by the remnant disk materials associated with the jet source
(please refer to Fig. 3 of Henney \& O'Dell 1999). Although this
does not directly affect emission from the central YSO, it appears
as red-displaced absorption in the extracted spectra of the
central source after background subtraction, which does not take
into account the disk obscuration. However, this feature is
specific to the source of HH 889 and was not detected toward
neither the HH 890 source (Li, Chu \& Gruendl 2005, sub.) nor the
only parallel jet sources in the immediate surroundings of
$\sigma$ Orionis (Reipurth et al. 1998; Andrews et al. 2004).

The spectrum of the HH 890 source is presented in Fig. 2, which
illustrates primarily moderate H$\alpha$, prominent [OIII]
emission and shows a late spectral type with a very red continuum
in the optical. The equivalent width of H$\alpha$ varies from 12.2
to 20.5~\AA~ in the two consecutive exposures, this leads to the
classification of the jet source as a CTTS based on predominantly
its H$\alpha$ emission. However, its high state of excitation as
revealed by its significant [OIII] emission lines strongly
suggests a fully ionized origin of the jet and at least the outer
layers of the relic disk in association, which will definitely
result in a rapid photodissipation of the system.

\section{Photoionization of the jet systems}

Two of the most massive exciting sources of the Rosette Nebula are
believed as the dominant sources of ionization and
photoevaporation of the jet system associated with HH 889. One is
HD 46223 with a spectral type of O4V(f), the hottest star in NGC
2244. It produces Lyman photons at a rate of 10$^{49.9}$ s$^{-1}$
(Panagia, 1973) and is located 277$"$ or 2.01 pc away from the jet
source if a distance of 1.5 kpc to Rosette is adopted (Dorland \&
Montmerle 1987). The other is HD 46150, an O5V star at a projected
distance of 433$"$ or 3.15 pc from the jet source. It resides at
the center of the HII region and produces ionizing photons at a
rate of 10$^{49.7}$ s$^{-1}$ (Panagia, 1973). The combined Lyman
continuum emission from these two stars has generated 1-2 orders
of magnitude more impact on the disk-jet system associated with HH
889 than that on similar jets discovered in the vicinity of
$\sigma$ Orionis (Reipurth et al. 1998) and the Trapezium stars
(Bally et al. 2000). The Rosette Nebula is therefore among the
most extreme environments in which irradiated jets are found.
Though immersed in photoionized medium, the existence of highly
collimated jets strongly suggests the existence of at least a
relic disk as a sustained feed to the surviving jet. In the case
of the Rosette irradiated jets, a photoevaporating disk with a
configuration resembling that of HH 527 in the Orion Nebula as
resolved by HST is expected (Bally et al. 2000). This could serve
as a schematic impression of the appearance of the disk jet
system, though in the case of HH 527, the jet is perhaps oriented
at a different direction as respect to the incident UV radiation,
is located in the outskirts of the Orion Nebula and has a low
excitation state. The physical configuration of the disk-jet
systems subject to photoelaboration induced disk-dissipation is
believed to be similar.

The HH 890 source is, however, found to be located in a less
disruptive UV radiation field, roughly an order of magnitude lower
than the HH 889 source faces. That is probably why this jet source
still show spectral properties resembling a CTTS, which put
constraints on the amount of circumstellar material still in
association. However, as stated above, its high ionization status
foretells the same fate for this jet system, although it may
sustain for a much longer period of time before the remnant disk
materials is finally stripped off the central protostar by the UV
photoevaporation and dissipation.

\section{Evidence for fast disk dissipation and a young stellar age?}

Being immersed in the fierce UV radiation field of Rosette, the
detection of optical jets associated with YSOs indicates either a
jet production timescale of as long as 1-2 Myrs, comparable to the
evolutionary age of the main cluster NGC 2244, or the YSOs must
have a much younger age and the cocoons associated with their
protostars had, in some way, been successfully shielded from the
strong ionization fields.

Infrared excessive emission is widely taken as an indicator of the
existence of circumstellar disks and, hence, youth of the young
stars in association.  Li (2005) investigated the near infrared
properties of the YSOs in the young open cluster NGC 2244 based on
the archived 2MASS database. Surprisingly, the jet-driving sources
in the Rosette do not show excess emission in the near infrared.
Instead, their colors are commensurate with those of WTTS, which
have spectral energy distributions indistinguishable from
main-sequence dwarves. The HH 890 source, though classified in
this paper as a CTTS primarily based on its intensive emission in
H$\alpha$, also shows infrared colors far different from that of
conventional CTTS. This suggests that the intense UV radiation
exerted by both the external massive stars and the internal
protostar has efficiently dissipated the bulk of the disk
associated with the YSO. Please refer to their positions on the
near infrared color-color and color-magnitude diagrams as
presented by Li (2005). This along with the fact that none of the
Rosette jet sources and their only rivals found near $\sigma$
Orionis were detected by the IRAS satellite suggests a lack of
circumstellar materials as compared to conventional YSOs driving
outflows. These are in agreement with the mass estimation of 0.006
M$\odot$, some two orders below the typical value of $\sim$0.1
M$\odot$ of the disks around CTTS, of the relic disk associated
with the HH 889 source by Li \& Rector (2004). Given the emerging
nature of the young open cluster with a turn-off age of 1.9 Myr
(Park \& Sung 2002), fast disk dissipation is suggested. Li \&
Rector (2004) suggest that this provides indirect observational
evidence for the formation of especially isolated BDs and free
floating giant planets as discovered in the best studied HII
region, the Orion Nebula (Zapatero Osorio et al. 2000), by UV
dissipation of unshielded protostellar cores. It can be very
important to our understanding of the formation of such
sub-stellar objects particularly in regions of massive star
formation.  Such processes of UV-dissipation could alternatively
impose strong effects on the formation and hence the search for
extra-solar planets around low-mass stars, the circumstellar disks
of which could otherwise be nice sites of terrestrial planet
formation. This introduces a viable solution to a long puzzle of
how WTTS were formed as a consequence of rapid CTTS evolution and
of the rapid dissipation of its circumstellar disk under
particular star forming conditions near massive OB stars or in
cluster environments. But could the lack of circumstellar
materials otherwise indicate an evolved status of the jet sources
in Rosette?

The spatial distribution of the two extreme jets in Rosette as
respect to the dozens of exciting OB stars of the spectacular HII
region is presented in Fig. 3. On which the relic shell structure
as indicated by the apparent congregation of excessive emission
sources in the near infrared (Li 2005) was superimposed, which
suggests the existence of a former working interface layer of the
HII region with its ambient molecular clouds.  Its projected
location of the HH 889 source near the relic arc could well
indicate evidence of a triggered origin of its formation in or
near the swept-up layer.  The HH 890 source has a similar radial
distance from the statistical center of NGC 2244 (Li 2005) and may
introduce a similar origin.  In this scenario, the jet sources
have a much younger age than that of the main cluster NGC 2244.
Molecular gas and dust associated with the shell may have played
an important role in protecting their circumstellar envelopes from
the strong UV ionization from the central OB stars. It is
therefore reasonable to infer that the jet sources have been
directly exposed to the photoionization field just recently.

\section{Conclusions}

We conclude that the Rosette sources driving jets may well
represent a transient phase of YSOs evolving rapidly from CTTS to
WTTS by fast UV dissipation of their circumstellar disks. This
results from abrupt loss of circumstellar materials by fierce
photoionization from nearby OB stars as observed in Rosette, or
rather happens in cluster environments due to strong interaction
between newly formed cluster members. This, however, poses strong
challenge to the traditional knowledge of jet formation, which
results from enhanced mass accretion and the release of angular
momentum of the jet-driving sources. Further detailed studies of
the jet systems bathed in photoionized environments may lead to a
final solution to the mechanisms of jet formation and collimation.
Observations with the Spitzer space telescope are strongly
suggested to further investigate the properties of the jet driving
sources in the mid-infrared.

\acknowledgements We would like to show our great appreciations to
an anonymous referee for the helpful comments on the paper. This
project is supported by the National Natural Science Foundation of
China through grant No.10503006.

{}
\bibliographystyle{aa}

\figcaption[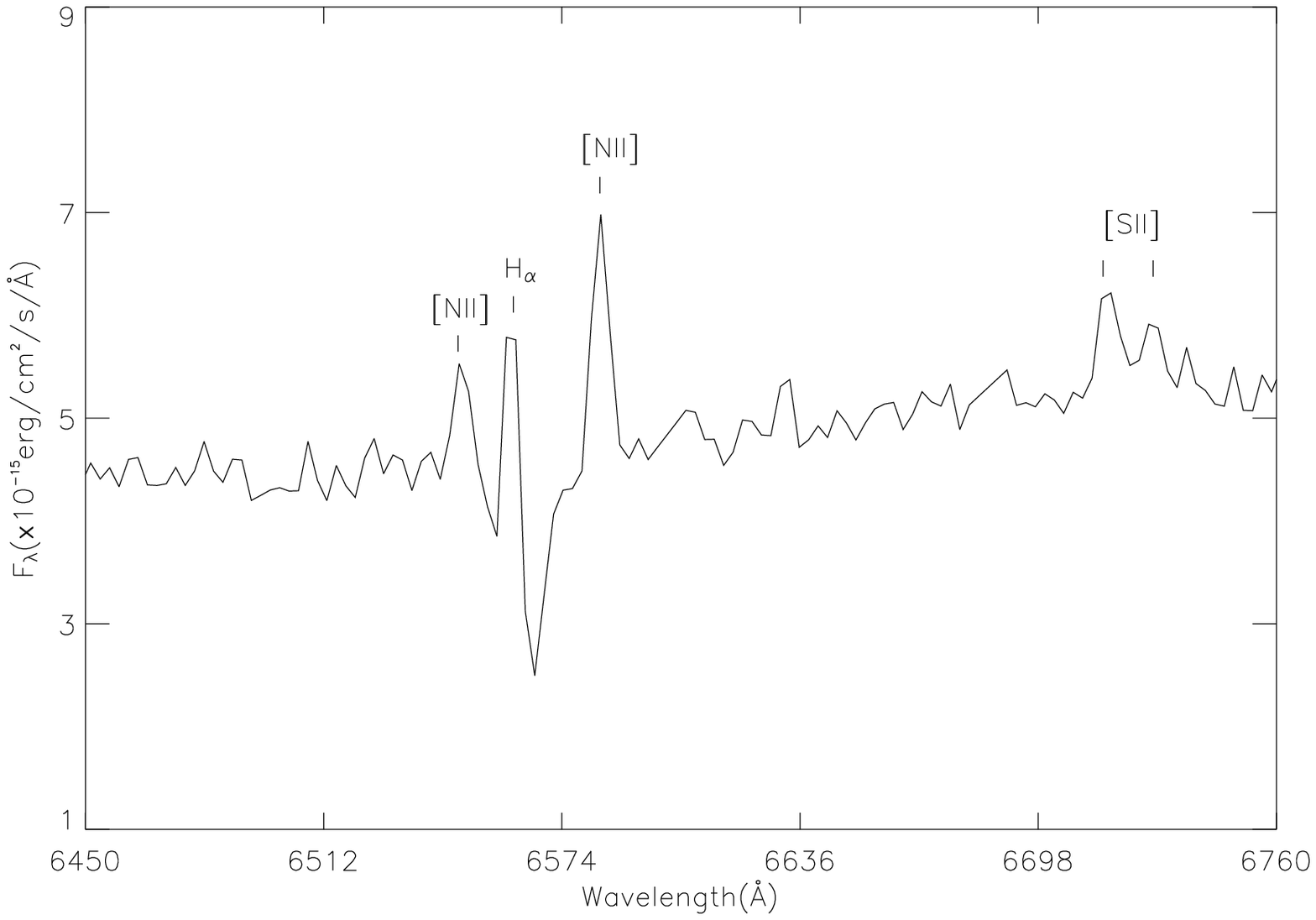]{Spectrum of the HH 889 source.}

\figcaption[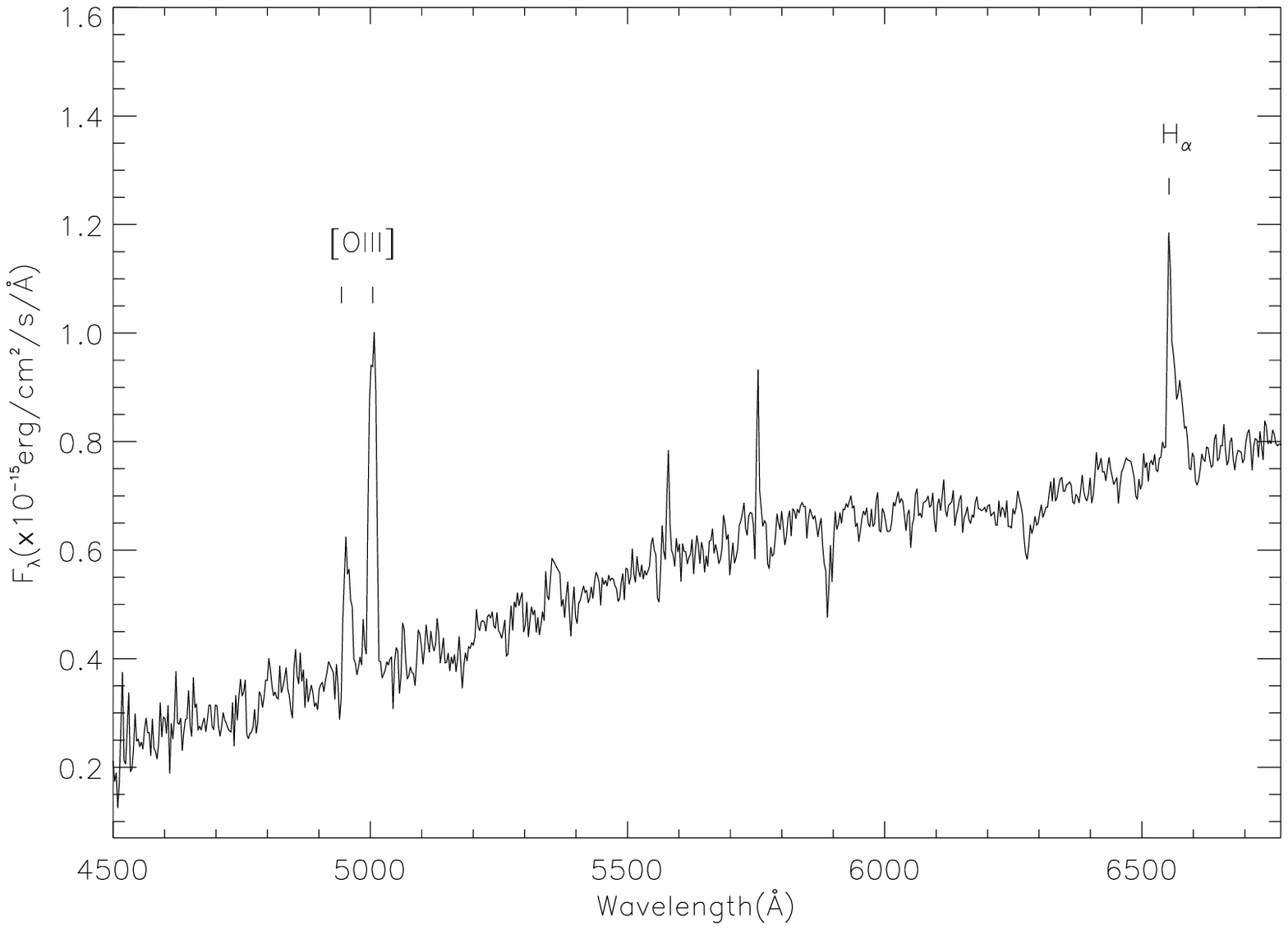]{Spectrum of the HH 890 source.}

\figcaption[f2.ps]{Spatial distribution of the Rosette HH jets as
respect to the ionizing massive OB stars that excavate the HII
region. The spectral type of each of the OB stars is also marked
on the DSS (Digital Sky Survey) R band image of Rosette. The
superposed circles mark the positions of the emerging cluster NGC
2244 (large) and its companion cluster (small) revealed by Li
(2005), the enclosed arc structure corresponds to a relic shell
harboring triggered star formation in the Rosette Nebula.}

\clearpage

\begin{figure}
\plotone{f1.ps} \centerline{f1.ps}
\end{figure}
\clearpage

\clearpage

\begin{figure}
\plotone{f2.ps} \centerline{f1.ps}
\end{figure}
\clearpage

\clearpage

\begin{figure}
\plotone{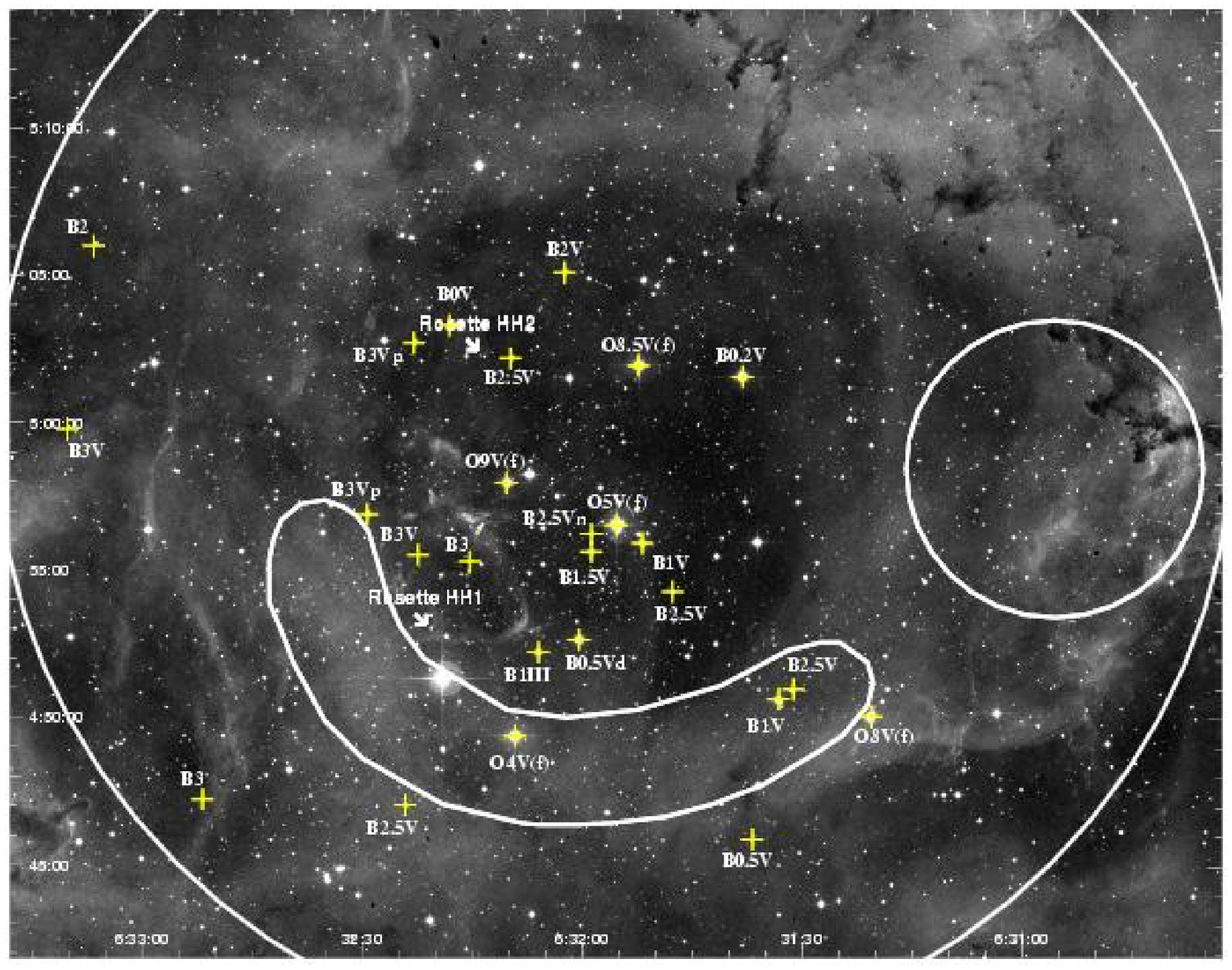} \centerline{f1.ps}
\end{figure}
\clearpage

\end{document}